\documentclass[pra,twocolumn]{revtex4}  %% REVTeX 4.0

\usepackage{graphicx}
\usepackage{amsmath,amssymb,amsfonts}
\usepackage{natbib}

\newcommand{\bd}{\begin{displaymath}}
\newcommand{\ed}{\end{displaymath}}
\newcommand{\be}{\begin{equation}}
\newcommand{\ee}{\end{equation}}
\newcommand{\ba}{\begin{eqnarray}}
\newcommand{\ea}{\end{eqnarray}}

\begin{document}

\title{Transmission properties in waveguides:
An optical streamline analysis}

\author{A. S. Sanz}
\email{asanz@iff.csic.es}
\affiliation{Instituto de F\'{\i}sica Fundamental (IFF--CSIC), Serrano
123, 28006 Madrid, Spain}

\author{J. Campos-Mart{\'\i}nez}
\email{jcm@iff.csic.es}
\affiliation{Instituto de F\'{\i}sica Fundamental (IFF--CSIC), Serrano
123, 28006 Madrid, Spain}

\author{S. Miret-Art\'es}
\affiliation{Instituto de F\'{\i}sica Fundamental (IFF--CSIC), Serrano
123, 28006 Madrid, Spain}

\begin{abstract}
A novel approach to study transmission through waveguides in terms of
optical streamlines is presented.
This theoretical framework combines the computational performance of
beam propagation methods with the possibility to monitor the passage of
light through the guiding medium by means of these sampler paths.
In this way, not only the optical flow along the waveguide can be
followed in detail, but also a fair estimate of the transmitted light
(intensity) can be accounted for by counting streamline arrivals with
starting points statistically distributed according to the input pulse.
Furthermore, this approach allows to elucidate the mechanism leading to
energy losses, namely a vortical dynamics, which can be advantageously
exploited in optimal waveguide design.

\vspace{.25cm}


\end{abstract}

%\ocis{000.0000, 999.9999.}% REPLACE WITH CORRECT OCIS CODES FOR YOUR ARTICLE
                          % NOTE: \ocis{} IS ALIASED TO \pacs{} BUT MUST
                          % FORMAT THE TERMS CORRECTLY FOR EACH JOURNAL

%\ocis{030.5260, 030.5290, 050.4865, 060.2310, 060.4230, 070.2580}

%030.5260   Photon counting
%030.5290   Photon statistics
%050.1960   Diffraction theory
%050.1970   Diffractive optics
%050.4865   Optical vortices
%060.2310   Fiber optics
%060.4230   Multiplexing
%070.2580   Paraxial wave optics
%070.7345   Wave propagation
%130.2790   Guided waves
%220.2560   Propagating methods
%230.1360   Beam splitters
%230.7020   Traveling-wave devices
%230.7370   Waveguides
%230.7380   Waveguides, channeled
%230.7390   Waveguides, planar
%230.7400   Waveguides, slab
%260.1960   Diffraction theory
%260.2160   Energy transfer

\maketitle

%%%%%%%%%%%%%%%%%%%%%%%%%%%%%%%%%%%%%%%%%%%%%%%%%%%%%%%%%%%%%%%%%%%%%%%

\section{Introduction}
\label{sec1}

Optical waveguide design is central to the development of efficient
devices with applications in optoelectronics and nanotechnologies
\cite{xiang:naturecomm:2011}.
More recently, it has also become relevant to quantum information
processing and quantum technologies based on photons
\cite{politi:science:2008,mataloni:PRL:2010}.
Very often these designs involve a complexity that arises from the
particular details of individual guiding elements as well as the
existence of a network of waveguides staged in series.
Thus, to ensure optimal guiding conditions, it is of interest
to develop efficient and flexible propagation and optimization
first-principle procedures, more specifically, implementing and
developing appropriate propagation methods to solve Maxwell's
equations.
These vector equations are highly demanding in terms of
computational effort if adequate balance in accuracy is required.

For some particular types of waveguides (e.g., optical
waveguides made of semiconductor materials), though, light propagation
can be properly accounted for by the scalar optics approximation, in
particular, the Helmholtz equation
\cite{banerjee-josaa:89,scarmozzino-rev:00,bornwolf-bk}.
In those cases where the waveguides do not bend too severely
and the ratio between the inside and outside refractive indexes is
relatively small, their study can be further simplified by appealing
to the paraxial or small-angle approximation.
The Helmholtz equation is then replaced by a simpler scalar wave
equation, namely the paraxial equation
\cite{banerjee-josaa:89,scarmozzino-rev:00}.
Of course, in more adverse cases the paraxial approximation can no
longer be used and other types of approximations can be considered,
such as the so-called wide angle equation \cite{pepe:AppOpt:2003},
but this goes beyond the scope of this work.

The analogy between optical and matter waves is a well-known subject
in optics
\cite{bornwolf-bk,buchdahl-bk,glogel:JOSA:1969,kahn:JOSA:1983},
which becomes more apparent when considering the isomorphism between
the paraxial equation and the time-dependent Schr\"odinger equation.
Hence, numerical methods available in quantum mechanics to propagate
wave packets
\cite{feit-fleck:JCompPhys:1982,feit-fleck:JCP:1983,leforestier:jcompphys:1991}
can also be used to study optimal waveguiding conditions for the
transmission of light pulses
\cite{pepe:JLightTech:1998,pepe:AppOpt:1999,longhi:PRE:2003}
---actually, in some cases, these methods came originally from optics,
which shows the fruitful interplay between these two branches of
physics---, where they are known as beam propagation methods (BPMs).

Now, apart from having efficient BPMs, one of the
key questions that arises in a natural way from the quantum-optical
analogy is whether some extra tools can be devised and used to analyze
and to better understand the transport properties inside waveguides.
In this regard, consider, for example, the analysis of diffraction
and interference with polarized light by means of photon trajectories
\cite{sanz:PhysScrPhoton:2009,sanz:AnnPhysPhoton:2010,sanz:JRLR:2010},
which are the direct optical analog of the quantum trajectories coming
from the Bohmian formulation of quantum mechanics
\cite{bohm:PR:1952-1,holland-bk}.
These studies, motivated by single-photon experiments
\cite{weis:AJP:2008,weis:PhysScr:2009,weis:EJP:2010}, allowed a better
understanding of the flux of electromagnetic energy in such contexts,
accounting for the formation of interference fringes or their
disappearance depending on the polarization of the interfering beams.
In fact, in spite of the theoretical nature of these studies, recently
photon trajectories have been inferred from experimental evidence
\cite{aephraim:Science:2011}, confirming the topology of those
previously reported theoretically \cite{sanz:AnnPhysPhoton:2010}.

In this work, an analogous concept to that of photon trajectory is
considered within the context of waveguiding and utilized to render
some light on how transport is conducted inside waveguides.
Unlike the photon trajectories mentioned above, here we take advantage
of the isomorphism between the paraxial equation and the time-dependent
Schr\"odinger equation, which allows us to readily extend the Bohmian
approach to the light transport accounted for by the former.
Since the light flux analyzed is time-independent, to avoid possible
connotations associated with the concept of photon trajectory (as
describing the time-evolution of particles rather than fluxes), the
concept of optical streamline will be used.
These streamlines are synthesized  ``on the fly'' in combination with
the BPM used to obtain the exact evolution of the pulse inside the
waveguide \cite{pepe:JLightTech:1998}.
As it is shown, the optical streamlines provide us with an alternative,
complementary insight to the standard wave approach, which allows us to
understand the flow of light inside waveguides by monitoring its way
through locally, i.e., just like a tracer particle allows us to
determine the flow of a classical fluid.
In this regard, this constitutes an interesting alternative tool to
analyze this kind of systems and their optical properties.
In particular, we show how optical streamlines provide us with detailed
dynamical (though stationary) information about guiding properties
(i.e., the transport of light) and the loss mechanism.
Although there are different designs depending on their particular
application, in order to properly introduce the language and
methodology, here we focus on the Y-junction structure,
which is widely used and appears as the main component in many other
more intricate optical devices.
More specifically, as a working model, the Y-junction geometry
developed by Langer and co-workers
\cite{coalson:JLightTech:1994,min-fiber:97} has been considered, which
is closely related to realistic waveguides of potential industrial
interest.

This work is organized as follows. In Section~\ref{sec2}, a brief
account on the theory behind both the paraxial equation and its
connection to Bohmian mechanics is presented.
Section~\ref{sec3} deals with the application of this theory to the
analysis of the light flow through Y-junction-type waveguides,
providing some numerical results.
Finally, in Section~\ref{sec4} the main
conclusions and future perspectives are summarized.

%%%%%%%%%%%%%%%%%%%%%%%%%%%%%%%%%%%%%%%%%%%%%%%%%%%%%%%%%%%%%%%%%%%%%%%
%%%%%%%%%%%%%%%%%%%%%%%%%%%%%%%%%%%%%%%%%%%%%%%%%%%%%%%%%%%%%%%%%%%%%%%

\section{Theory}
\label{sec2}

Consider the optical axis of the waveguide is oriented along the
$z$-direction, while its transversal section is parallel to the
$XY$-plane.
Moreover, the passage of light through the waveguide is describable
in terms of a time-harmonic electromagnetic field, which allows us to
simplify the treatment by considering a stationary or time-independent
scalar field $\Phi({\bf r})$ \cite{bornwolf-bk}.
This implies $\Phi({\bf r})$ satisfies Helmholtz's equation,
\begin{equation}
 \nabla^2 \Phi + n^2 k^2 \Phi = 0 ,
 \label{1}
\end{equation}
plus the corresponding boundary conditions imposed by the shape and
refractive properties of the waveguide.
In Eq.~(\ref{1}), $n \equiv n({\bf r})$ is the position-dependent
refractive index inside the waveguide and $k = 2\pi/\lambda$,
with $\lambda$ being the light wavelength in vacuum.

Taking into account this configuration for the waveguide, in the
small-angle or paraxial approximation, the scalar field $\Phi$ can be
approximated by a plane wave along the $z$-direction modulated by a
certain complex-valued amplitude \cite{bornwolf-bk}, i.e.,
\begin{equation}
 \Phi({\bf r}) = \phi({\bf r}) e^{ik_z z} .
 \label{par1}
\end{equation}
In other words, this means that, for this type of nonuniform waveguides,
fast oscillations can be separated from slower ones.
Accordingly, when moving inside the bulk, this splitting allows us to
write $k_z = \beta_0 = kn_0$, where $n_0$ is the {\it bulk index}.
Substituting (\ref{par1}) into Helmholtz's equation (\ref{1}), the
latter can be recast as
\begin{equation}
 2ik_z\ \frac{\partial \phi}{\partial z}
  + \frac{\partial^2 \phi}{\partial z^2}
  = - \nabla_\perp^2 \phi + (k_z^2 - k^2n^2) \phi .
 \label{part2}
\end{equation}
Here, $\nabla_\perp^2$ denotes the transverse Laplacian, i.e., the
Laplacian on the subspace perpendicular to the propagation direction,
$\nabla_\perp^2 = \partial^2/\partial x^2 + \partial^2/\partial y^2$.
Apart from the paraxial approximation, if we also assume the slowly
varying envelope approximation holds, i.e., the envelope $\phi$ varies
slowly in space compared to $2\pi/k_z$, the highest-order derivative,
$\partial^2 \phi/\partial z^2$, can be neglected in Eq.~(\ref{part2}),
which can be expressed as
\be
 2i\beta_0\ \frac{\partial \phi}{\partial z} =
  - \nabla^2_\perp \phi + \left( \beta_0^2 - k^2n^2 \right) \phi .
 \label{4}
\ee
As it can be readily noticed, this equation is isomorphic to the
time-dependent Schr\"odinger equation, except for the evolution,
which is not in time, but in terms of the $z$-coordinate (within this
context the space $z$-coordinate acts as the ``evolution'' parameter).
From now on, the transversal coordinates $(x,y)$ will be denoted
collectively by means of the vector ${\bf R}$.
This quantum-optical analogy becomes more apparent if Eq.~(\ref{4})
is recast as
\be
 i\ \frac{\partial \phi}{\partial z}
  = \left[ -\frac{1}{2kn_0}\ \nabla_\perp^2 +
    V({\bf r}) \right] \phi ,
 \label{7}
\ee
where
\be
 V({\bf r}) = \frac{k}{2n_0} \left[ n_0^2 - n^2({\bf r}) \right]
 \label{10}
\ee
is an effective  potential function accounting for the waveguide
refractive profile and $\mu \equiv \beta_0 = k n_0$ plays the role
of an ``optical mass'' \cite{tien:RMP:1977,tien:radiosci:1981}.
Here, in particular, we have chosen an optical effective mass to be
$\mu = 2\beta_0$ in order to enhance the dynamical effects inside the
guide.
Nevertheless, it is important to stress that, while Schr\"odinger's
equation describes a true time-evolution, Eq.~(\ref{7}) only describes
how the stationary, three-dimensional scalar field $\phi$ (and,
therefore, the light intensity) distributes along the waveguide.
The fact that this field behaves differently along the $z$-coordinate
than along the transversal section is used conveniently from a
computational viewpoint to solve the corresponding three-dimensional
wave equation, but the field itself is always stationary.

Following with the quantum-optical analogy, consider $\phi$ is now
expressed in polar form,
\be
 \phi({\bf R},z) = \rho^{1/2}({\bf R},z) e^{iS({\bf R},z)} ,
 \label{8}
\ee
in order to establish a parallelism with Bohmian mechanics
\cite{bohm:PR:1952-1,holland-bk}, also known as quantum hydrodynamics
\cite{madelung:ZPhys:1926,birula-bk}.
Within this formulation of quantum mechanics, $\rho$ and $S$ denote,
respectively, the probability density and phase of the wave function
describing a quantum particle of mass $m$, both being real-valued
functions.
In the present context, though, these quantities are related,
respectively, to the (stationary) energy density inside the waveguide
and its flux.
After substitution of (\ref{8}) into the optical Schr\"odinger-like
equation (\ref{7}), we obtain a system of two real coupled equations,
\begin{eqnarray}
 \frac{\partial \rho}{\partial z} & + & \nabla \cdot {\bf J} = 0 ,
 \label{e3} \\
 \frac{\partial S}{\partial z} & + &
  \frac{\left( \nabla_\perp S \right)^2}{2\mu} + V_{\rm eff} = 0 .
 \label{e4}
\end{eqnarray}
Equation~(\ref{e3}) is a continuity equation, which describes how the
energy is distributed throughout the waveguide.
However, given the parametrization in terms of $z$, this equation can
also be interpreted as accounting for the connection between the
energy distribution $\rho = \phi^*\phi$ along the transversal section
at a certain value of $z$ with its flux through this section, ${\bf J}$.
This flux can be expressed as ${\bf J} = \rho {\bf v}$, i.e., in terms
of a local velocity field,
\begin{equation}
 {\bf v} = \frac{\nabla_\perp S}{\mu}
   = \frac{\hbar}{2\mu i} \left(
   \frac{\phi^*\nabla_\perp \phi - \phi \nabla_\perp \phi^*}
    {\phi^* \phi} \right) .
 \label{e5}
\end{equation}
Assuming ${\bf v} = {\bf R}' = d{\bf R}/dz$ and then integrating for
$x$ and $y$ along the time-like $z$-coordinate with different initial
conditions, one obtains a series of optical streamlines.
These streamlines allow us to visualize how energy flows inside the
guide at a local level.
That is, we can connect causally a particular point of the input pulse
with another one of the output one in a non-ambiguous fashion ---unlike
classical rays, the optical streamlines evolve in space in accordance
with the wave-like behavior of the field $\phi$.
In quantum mechanics the corresponding quantum
streamlines monitor the flow described by the quantum (probabilistic)
fluid and have been used earlier in the literature to describe
different physical systems
\cite{birula:PRD:1971,hirsch:JCP:1974-1,hirsch:JCP:1974-2,hirsch:JCP:1976-1,hirsch:JCP:1976-2}
as well as to devise new numerical propagation schemes \cite{wyatt-bk}.

In order to better understand the meaning of two points being causally
connected, we need to focus on Eq.~(\ref{e4}).
This equation describes the evolution (in terms of the $z$-coordinate)
of the phase field $S$ along the waveguide and presents the form
of a quantum Hamilton-Jacobi equation \cite{bohm:PR:1952-1,holland-bk}.
In it, the effective potential $V_{\rm eff}$ consists of two
contributions, the potential $V$ defined above [see Eq.~(\ref{10})]
and a kind of ``optical'' potential,
\begin{eqnarray}
 Q \equiv - \frac{1}{2\mu} \frac{\nabla^2 \rho^{1/2}}{\rho^{1/2}}
 = \frac{1}{4\mu}
   \left[ \frac{1}{2} \left( \frac{\nabla \rho}{\rho} \right)^2
   - \frac{\nabla^2 \rho}{\rho} \right] ,
 \label{e6}
\end{eqnarray}
which is analogous to the {\it quantum potential} of Bohmian mechanics.
As in the classical version of the Hamilton-Jacobi mechanics, solutions
of Eq.~(\ref{e4}) define trajectories as the perpendicular streamlines
to surfaces of constant phase $S$, which here translates into the
equation of motion (\ref{e5}).
The coupling between (\ref{e3}) and (\ref{e4}) through $Q$ (or,
equivalently, $\rho$ and its space derivatives) is the reason why the
wave formulation for waveguides differs from its geometric counterpart.
The guidance condition (\ref{e5}) thus allows to properly include
typical wave-like features like interference and diffraction, for
example, in the topology of the streamlines, which does not happen,
however, in the rays one obtains from geometric optics.

%%%%%%%%%%%%%%%%%%%%%%%%%%%%%%%%%%%%%%%%%%%%%%%%%%%%%%%%%%%%%%%%%%%%%%%
%%%%%%%%%%%%%%%%%%%%%%%%%%%%%%%%%%%%%%%%%%%%%%%%%%%%%%%%%%%%%%%%%%%%%%%

\section{Numerical results}
\label{sec3}

%%%%%%%%%%%%%%%%%%%%%%%%%%%%%%%%%%%%%%%%%%%%%%%%%%%%%%%%%%%%%%%%%%%%%%%

\subsection{The waveguide design}
\label{sec31}

Consider a waveguide that transports a light beam injected at $z = 0$
to a given distance $z = L$.
Apart from its particular geometry, the transport properties of this
waveguide are characterized by the bulk refractive index ($n_0$),
the guide refractive index ($n_1$), and the cladding refractive index
($n_2$).
Physically, these refractive indexes can be associated with an
effective potential function that accounts for the waveguide
refractive profile (the profile along the transversal direction),
as seen in the previous section.
Inside the waveguide this effective potential or refractive profile
reads as
\be
 V = \frac{k}{2n_0} \left( n_0^2 - n_1^2 \right) ,
 \label{12}
\ee
while outside it is given by
\be
 V = \frac{k}{2n_0} \left( n_0^2 - n_2^2 \right) .
 \label{13}
\ee
In order to simplify our study, but without loss of generality, we are
also going to assume that the electric field depends only on one of the
two transversal coordinates, say $x$.
This is the condition that determines the confinement of light inside
the waveguide \cite{bornwolf-bk}.
The other direction, $y$, will be less relevant, only adding some
attenuation when moving further away inside the bulk.
Thus, in the present case, the distribution of light inside the
waveguide is accounted for by a stationary (time-independent) scalar
field $\phi(x,z)$.
Alternatively, taking into account the wave picture described in the
previous section, $\phi$ can also be interpreted as describing the
evolution of a pulse of light distributed along the $x$ (space-like)
direction as the value of $z$ (time-like) coordinate increases.

\begin{figure}[!h]
 \begin{center}
  \includegraphics[width=8cm]{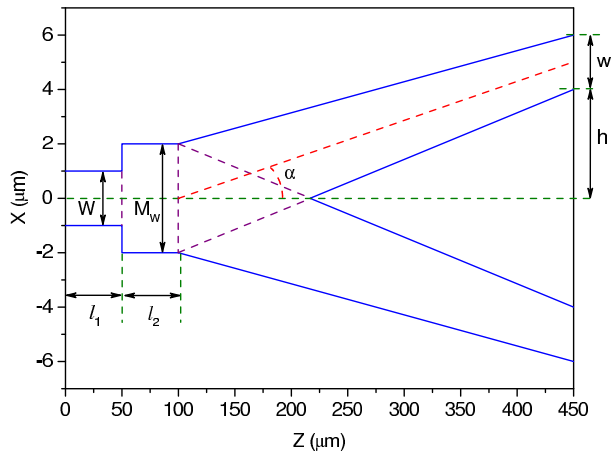}
  \caption{Geometrical layout of the Y-junction design considered here,
   where the significant (geometrical) parameters are also displayed
   (see Table~\ref{table1}).}
  \label{fig1}
 \end{center}
\end{figure}

The configuration of the waveguide along the $z$-direction is as
displayed in Fig.~\ref{fig1}.
It is a Y-junction which utilizes a 1$\times$2 multimode interference
(MMI) device with a lowered refractive index ($n_w$) in the junction
or wedge region, as proposed by Langer {\it et al.}\ \cite{min-fiber:97}
(here, though, we have assumed the wedge refractive index to be equal
to the guide one).
This wedge reduces the loses over a rather large range of
Y-junction angles.
Details on the numerical values for the geometric and refractive index
parameters considered here are given in Table~\ref{table1}.

%%%%%%%%%%%%%%%%%%%%%%%%%%%%%%%%%%%%%%%%%%%%%%%%%%%%%%%%%%%%%%%%%%%%%%%

\subsection{Numerical methodology}
\label{sec32}

In this work, the Schr\"odinger-like form of the paraxial equation,
Eq.~(\ref{7}), has been solved by taking advantage of the
computational machinery developed in standard quantum wave-packet
propagations to solve the time-dependent Schr\"odinger equation.
More specifically, we have considered the split-operator technique
combined with the fast Fourier transform method
\cite{feit-fleck:JCompPhys:1982,feit-fleck:JCP:1983,leforestier:jcompphys:1991}.
As it has been shown \cite{pepe:JLightTech:1998,pepe:AppOpt:1999},
this type of BPM results very efficient computationally, particularly
in optimal control scenarios.

\begin{figure*}
 \centerline{\includegraphics[height=4.8cm]{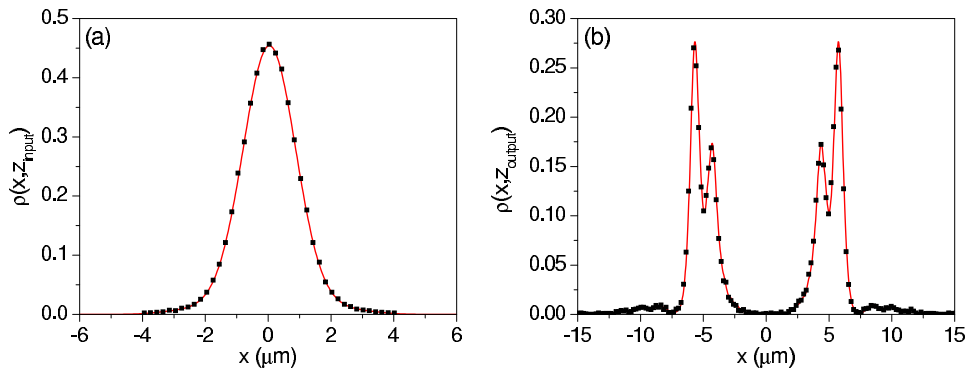}}
  \caption{Initial (a) and final (b) intensity transmitted through the
   waveguide.
   In each panel, black squares denote the optical streamline
   calculation (histogram), while the red solid accounts for the value
   obtained from the standard optical calculation (paraxial equation).
   The parameters used in these calculations are given in
   Table~\ref{table1}.}
  \label{fig2}
\end{figure*}

\begin{table}[!b]
 \caption{Values of the different parameters considered in the design
  of the waveguide used in the calculations shown here (see text for
  details).}
 \label{table1}
 \tabcolsep=20pt
 \begin{center}
  \begin{tabular}{cc}
  \hline\hline
   Parameter &  \\
   \hline
   $\lambda$ ($\mu$m) &  1  \\
   $W$ ($\mu$m)      &   2  \\
   $M_W$ ($\mu$m)    &   4  \\
   $\ell_1$ ($\mu$m) &  50 \\
   $\ell_2$ ($\mu$m) & 100  \\
   $h$ ($\mu$m)      &   4  \\
   $w$ ($\mu$m)      &   2  \\
   $\alpha$          & $\sim$0.82$^\circ$  \\
   $n_0$   & 3.558  \\
   $n_1$   & 3.568  \\
   $n_2$   & 3.558  \\
   $n_w$   & 3.568  \\
  \hline\hline
  \end{tabular}
 \end{center}
\end{table}

As mentioned above, the purpose of this work is to combine the
computational performance of BPMs with the physical insight provided
by the optical streamlines.
Therefore, regarding the calculation of the optical streamlines, we
have chosen to solve Eq.~(\ref{e5}) ``on the fly'', i.e., once $\phi$
has been evolved from $z$ to $z + \delta z$, the streamline is
updated, from $x(z)$ to $x(z+\delta z)$, by means of a fourth-order
Runge-Kutta integrator fed with the field $\phi$, according to the
last equality in (\ref{e5}).
This is a method that has been previously applied to the study of
matter-wave diffraction by periodic gratings \cite{sanz:JPCM:2002}.
Nevertheless, from a merely computational point of view, one could
also consider any of the so-called quantum trajectory methods
\cite{wyatt-bk}, based on solving also ``on the fly'' Eqs.~(\ref{e3}),
(\ref{e4}) and (\ref{e5}), which would skip solving the paraxial
equation.

Once the method is set up, another important element when dealing with
optical streamlines is that, in order to obtain a good representation
of the pulse at each value of the $z$-coordinate, one needs to have an
optimal sampling of the initial pulse, namely $\rho(x,0)$.
Since one of the purposes here is to show that the evolution of the
pulse along the waveguide can be followed by looking at the density of
streamlines rather than monitoring $\phi$ itself, we have not optimized
the number of streamlines considered.
Thus, a total of 11,256 initial positions $x_0 = x(z=0)$ have been
randomly generated according to the weight $\rho(x,0)$
\cite{sanz:CP:2011}.
This sampling has been made in such a way that the value of
$\rho(x_0,0)$ was confined between the maximum of the intensity
distribution (here, at $x=0$) and 10$^{-3}$ times this value, i.e.,
$\rho(x=0,0) \ge \rho(x_0,0) \ge 10^{-3}\times \rho(x=0,0)$, for all
$x_0$ [see Fig.~\ref{fig2}(a)].
These initial conditions have then been propagated and, at each value
of $z$, a histogram was built up with them to compare with the
intensity obtained from the paraxial equation.
This step in the calculation is crucial in order to obtain a fair
reproduction of the optical result.
We have observed that the calculation of the streamlines ``on the
fly'', as described above, does not decrease much the efficiency of
the BPM considered when computing a low number of sampler streamlines,
i.e., streamlines calculated to determine the flux of light without
caring much about having a good sampling.
Otherwise, the time of computation increases {\it strictly} linearly
with the number of streamlines, as in any standard trajectory
calculation.

%%%%%%%%%%%%%%%%%%%%%%%%%%%%%%%%%%%%%%%%%%%%%%%%%%%%%%%%%%%%%%%%%%%%%%%

\subsection{Analysis of the results}
\label{sec33}

\begin{figure}[!b]
 \begin{center}
  \includegraphics[width=8cm]{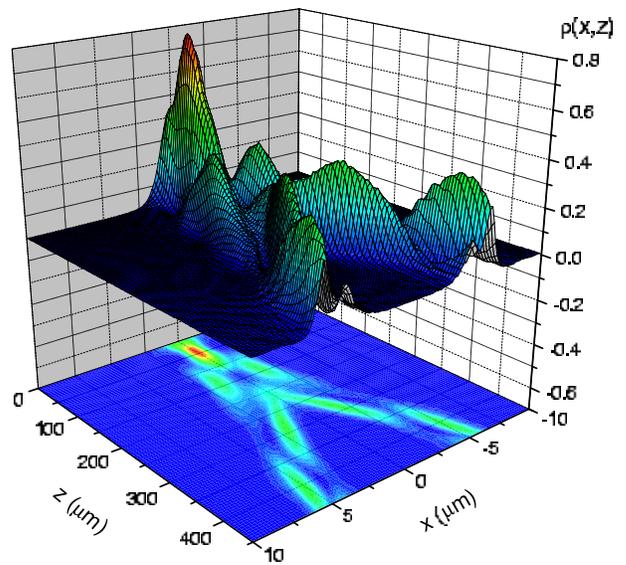}
  \caption{Three-dimensional representation showing the evolution
   of the initial pulse $\rho(x,0)$ as it propagates (along the
   $z$-coordinate) throughout the waveguide (below, the corresponding
   contour-plot is also displayed).
   In particular, $\rho(x,z)$ has been obtained as a histogram-like
   distribution of optical streamlines at each $z$ value, which
   corresponds fairly well with the representation one would find
   by standard wave propagation methods (see text for details).
   The parameters used in these calculations are given in
   Table~\ref{table1}.}
  \label{fig3}
 \end{center}
\end{figure}

In Fig.~\ref{fig2}(a), it is observed how the histogram fairly
reproduces the profile (along the waveguide transversal section)
of the input pulse at $z = 0$~$\mu$m.
The same is also found for larger values of the $z$-coordinate, at
$z = 450$~$\mu$m, as seen in Fig.~\ref{fig2}(b).
In general, this behavior can be found at any value of $z$, which
allows us to study the transmission properties along the waveguide
by only considering the propagation of the histogram, as shown in
Fig.~\ref{fig3}.
In this plot, the intensity reaches a
maximum as soon as it leaves the 1$\times$2 MMI device and then
splits up into two identical fluxes, each one displaying a series
of bounces inside its corresponding waveguide as they propagate.
These bounces can be somehow related to the bounces that a light ray
would undergo according to a geometrical optical viewpoint (the
ray propagates along a straight line until it reaches the boundary
of the waveguide, which defines its turning point in accordance with
the laws of reflection and refraction).

\begin{figure*}
 \centerline{\includegraphics[height=4.8cm]{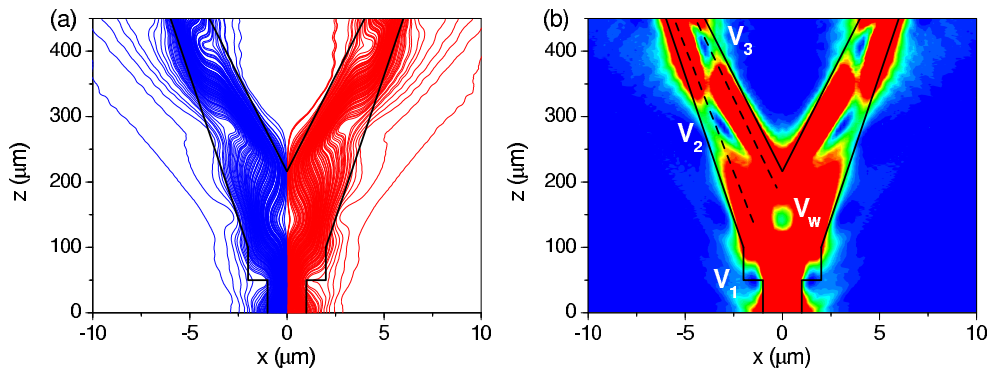}}
  \caption{Optical streamlines (a) and contour-plot of the evolution of the
   histogram (b) illustrating the flow of light throughout the waveguide.
   In both figures, the straight black lines mark the waveguide boundaries
   (see Fig.~\ref{fig1}).
   In part (a), blue and red lines are used to emphasize the fact optical
   streamlines also satisfy the non-crossing rule that characterizes both
   quantum \cite{sanz:JPA:2008} and optical fluxes
   \cite{sanz:AnnPhysPhoton:2010}.
   In part (b), the maximum value of the contours is set up to 0.1 in order
   to appreciate the outgoing light flow.
   White labels correspond to different vortices that appear during the
   evolution of the flow (see text for details).}
  \label{fig4}
\end{figure*}

The analogy with geometrical rays can be extended by means of the
optical streamlines here defined.
In Fig.~\ref{fig4}, a set of 200 of such streamlines, chosen (in
proportion) from among the full set run, are plotted.
As it can be noticed, optical streamlines behave as geometrical rays,
undergoing a series of bounces and reflection at the boundaries of the
waveguide, although their flow is more laminar, like tracer particles
moving on a fluid.
This is a consequence of the fact that they follow a wave and therefore
they have to accommodate to the features undergone by such a wave, such
as diffraction, interference or refraction.
In connection with this property, also note that they satisfy the
non-crossing property of Bohmian trajectories \cite{sanz:JPA:2008}:
two streamlines cannot pass through the same position $x$ for
the same value of the evolution coordinate, $z$.
This allows us to establish a clear difference between streamlines
evolving along the two branches of the Y-junction.
In the case considered here, this might seem to be not so important,
for the evolution along $z$ is symmetric with respect to $x=0$.
However, notice that as soon as a slight asymmetry between pathways
would appear (different refractive index, length, shape, etc.), the
splitting would also be different.
In such a case, the optical streamlines will provide us with
unambiguous information on which specific portion of the input
signal goes through each branch.
Thus, this type of information, which cannot be disentangled in
wave-based treatments by their intrinsically global nature, can be
of interest in applications based on waveguide technologies, such as
Mach-Zehnder interferometry or other optical interference devices.
Actually, even in the present case, this property results useful,
because it allows us to causally connect \cite{sanz:arxiv:2011} a
certain feature from the output signal at $z=450$~$\mu$m with a
particular region of the input, at $z=0$.
Later on this information can be used to make the sampling only over
a certain subset of initial conditions (e.g., those that does not
lead to losses at $z=450$~$\mu$m).

Apart from the properties above, an additional interesting feature
enabled by an analysis based on optical streamlines is the
detection of energy losses through the waveguide boundaries.
Detection of such losses by standard means, i.e., by only studying
the evolution of $\phi(x,z)$, the intensity $\rho(x,z)$ needs
to be integrated over the waveguide cross-section (at a given $z$
value).
From a more graphical viewpoint, the intensity should be
enlarged until the amount of intensity that is flowing outside is
perceived [see Fig.~\ref{fig4}(b)]. For example, in
Fig.~\ref{fig2}(b) the presence of two ``wings'' lying out of
the waveguide boundaries is clearly seen, which indicate the
leakage of light from the waveguide. However, they are neither
perceived in the contour-plot of Fig.~\ref{fig3}, nor the
continuous outwards flow of energy (or how and when this flow
happens). A more convenient and insightful way to understand this
flow arises by observing the topology of the optical streamlines,
as it is apparent from Fig.~\ref{fig4}(a). Although the
outermost streamlines are associated with very low intensity (that
is precisely the reason why they appear so sparse in the figure),
it is readily noticed how the flow is outwards. Actually, whenever
a node of the intensity appears, it pushes away the streamlines,
as can be seen at $z \approx 150$~$\mu$m, 275~$\mu$m or 450~$\mu$m
(and inwardly at $z \approx 400$~$\mu$m). In this sense, optical
streamlines acquire the role of optical tracers of the energy flow
inside the waveguide, allowing us to monitor it at each step. This
property thus results very useful in the optimal design of
waveguides and the control of light inside them, for it gives us a
straightforward method to test losses and gains.

At a more detailed level of description, the optical streamlines
provide us with a better understanding on the dynamics that is taking
place inside the waveguide, such as the different features which one
can observe later on in the modes or energy losses.
In this regard, consider first the high intensity peak observed at the
entrance of the second section of the 1$\times$2 MMI (see
Fig.~\ref{fig3}), which later on gives rise to a node or vortex
[$V_w$, in Fig.~\ref{fig4}(b)] in the wedge center.
By inspecting the optical streamlines [see Fig.~\ref{fig4}(a)], one
readily notices how most of them quickly undergo an inwards motion as
$z$ increases, just before $V_w$ becomes incipient at the position
where the second section of the 1$\times$2 MMI starts (regarding the
streamlines that surmount the vortex through the opposite side; see
comments below).
By construction, the tight squeezing of the swarm of streamlines takes
place just at the entrance of the second section of the 1$\times$2 MMI,
from which it undergoes a fast expansion.
This expansion causes the splitting of the beam into two distinctive
halves (partial beams), each one channeled into one of the branch of
the Y-junction and, moreover, a vortex in between.
It is interesting to compare this behavior to that of a classical
fluid which is released from a pipe, thus giving rise to channeled
streams and hollows or low density regions in between.

Now, consider only one of the branches of the waveguide, for example,
the left-hand side one. From a simple inspection of Fig.~\ref{fig4}(b),
where the contours refer to low intensity values (up to 10\% of its
maximum, at the entrance of second section of the 1$\times$2 MMI; see
Fig.~\ref{fig3}), one observes that the flow of energy appears and
disappears alternatively along two pathways as $z$ increases (see
black dashed lines).
This is an effect of the modes generated within the branches of the
waveguide that can be easily explained in terms of a simple model.
Consider that two kind of wave packets travel
along each pathway.
They are being populated alternatively as they evolve along $z$, i.e.,
when one starts to gain population, the other losses it, just as in a
simple two-level system,
\be
 \phi(x,z) = \cos (\omega z) \varphi_1(x)
  + \sin (\omega z) \varphi_2(x) .
\ee
According to this simple model one can easily explain that, for
example, near the vortex $V_2$ [see Fig.~\ref{fig4}(b)], one of the
pathways starts to get depopulated very quickly, while the other
reaches its maximum (see also Fig.~\ref{fig3}).
The same behavior is again observed when approaching $V_3$.
However, although this two-level model describes the features in the
population changes as the optical modes travel through the
waveguide branches, no information is provided about the
population transfer between pathways or energy losses. These are
processes suitable to be explained by means of the associated
optical streamlines, for they provide us with precise information
about the energy flux and its evolution along the waveguide, as
seen above. Thus, as it can be seen as the mode approaches $V_2$
[see Fig.~\ref{fig4}(a)], the streamlines start to move
smoothly rightwards, from one pathway to the other, until their
number has decreased considerably in the first pathway. Near $V_2$
there are three options for those streamlines still remaining
along the first pathway: (1) some of them undergo a sudden turn
and will enter into the second pathway stream, (2) some will
slightly bend and will continue along the first pathway beyond the
vortex, and (3) another group leave the waveguide after the
bending, thus giving rise to the losses (eventually, streamlines
from group~2 will also scape). This behavior can be observed not
only around $V_2$, but around any other external vortex ($V_1$,
$V_3$ and the homologous vortices along the second pathway), i.e.,
any vortex facing the boundaries of the waveguide. The goal of
waveguide optimal design is to avoid losses and therefore it can
highly benefit from this optical streamline representations.

%%%%%%%%%%%%%%%%%%%%%%%%%%%%%%%%%%%%%%%%%%%%%%%%%%%%%%%%%%%%%%%%%%%%%%%
%%%%%%%%%%%%%%%%%%%%%%%%%%%%%%%%%%%%%%%%%%%%%%%%%%%%%%%%%%%%%%%%%%%%%%%

\section{Summary and future perspectives}
 \label{sec4}

The purpose here was to introduce the concept of optical streamline in
the literature of waveguides as an analytical tool, in combination with
efficient BPMs to carry out their propagation.
To carry out our hydrodynamical analysis, we have considered as a
working model a Y-junction with a geometry closely related to realistic
waveguides of potential industrial interest, originally proposed some
year ago within the broad family of Y-junctions \cite{min-fiber:97}.
The particularity of these devices is that their design seeks for a
minimum loss of the input while having a clean splitting of the signal,
at the same time.
Thus, given the interest and relevance of these designs in modern
technologies, to have at our disposal additional tools which can help
us to design, study, explain and understand such systems is of great
importance.
In this regard, here we have mainly focused on determining how the
modes travel through the wave as well as on elucidating the
energy/light loss mechanism in waveguides.
These two features can be efficiently controlled by means of an
appropriate design \cite{coalson:JLightTech:1994,pepe:AppOpt:1999},
which is precisely the scope of the waveguide optimal design techniques
and the eventual target of our optical streamline analysis, where it
can be of much interest and help.
This is actually a natural and important extension of the present work,
which is currently under development.
Nonetheless, we also like to stress the general usefulness of the
proposed new tool, which might be of relevance in other more complex
scenarios where interference or merging of different
signals could be more efficiently studied (e.g., Mach-Zehnder
interferometers \cite{pepe:AppOpt:1999}), as well as other domains
going beyond the paraxial approximation (e.g., the wide-angle equation
\cite{pepe:AppOpt:2003}).

%%%%%%%%%%%%%%%%%%%%%%%%%%%%%%%%%%%%%%%%%%%%%%%%%%%%%%%%%%%%%%%%%%%%%%%
%%%%%%%%%%%%%%%%%%%%%%%%%%%%%%%%%%%%%%%%%%%%%%%%%%%%%%%%%%%%%%%%%%%%%%%

\section*{Acknowledgments}

This work has been supported by the Ministerio de Econom{\'\i}a y
Competitividad (Spain) under Projects FIS2010-22064-C02-02,
FIS2010-18132 and FIS2010-22082, as well as by the COST Action MP1006
({\it Fundamental Problems in Quantum Physics}).
A. S. Sanz would also like to thank the Ministerio de Econom{\'\i}a y
Competitividad for a ``Ram\'on y Cajal'' Research Fellowship.

%%%%%%%%%%%%%%%%%%%%%%%%%%%%%%%%%%%%%%%%%%%%%%%%%%%%%%%%%%%%%%%%%%%%%%%
%%%%%%%%%%%%%%%%%%%%%%%%%%%%%%%%%%%%%%%%%%%%%%%%%%%%%%%%%%%%%%%%%%%%%%%

%\bibliography{references}

%\end{document}

\end{document}